\documentclass[a4paper,10pt]{article}
\usepackage[margin=3cm]{geometry}
\usepackage{graphicx}
\usepackage{dsfont,amsmath,amssymb,color,comment}

\newcommand{\be}{\begin{equation}} \newcommand{\ee}{\end{equation}}

\newcommand{\ben}{\begin{enumerate}} \newcommand{\een}{\end{enumerate}}
\newcommand{\bc}{\begin{center}} \newcommand{\ec}{\end{center}}
\newcommand{\bi}{\begin{itemize}} \newcommand{\ei}{\end{itemize}}

\newcommand \bea {\begin{eqnarray} \nonumber }
\newcommand \eea {\end{eqnarray}}

\begin{document}
 
\title{A Fokker-Planck description for the queue dynamics of large tick stocks}

\author{A. Gar\`eche\,$^{\textrm{a,b}}$, G. Disdier\,$^{\textrm{a,c}}$,
\\ J. Kockelkoren\,$^{\textrm{a}}$, J.-P. Bouchaud\,$^{\textrm{a}}$}

\date{\today}
\maketitle
\small
\begin{center}
$^\textrm{a}$~\emph{Capital Fund Management, 23 rue de l'Universit\'e, Paris, France}\\
$^\textrm{b}$ now at:~\emph{Marshall Wace LLP, The Adelphi, 1/11 John Adam Street \\
London WC2N 6HT, UK}\\
$^\textrm{c}$ now MFE Candidate,~\emph{University of California, Berkeley, USA}\\
\end{center}
\normalsize

\vspace{0.3cm}

\begin{abstract}
Motivated by empirical data, we develop a statistical description of the queue dynamics for large tick assets based on a two-dimensional Fokker-Planck (diffusion) equation, that
explicitly includes {\it state dependence}, i.e. the fact that the drift and diffusion depends on the volume present on both sides of the spread. ``Jump'' events, corresponding 
to sudden changes of the best limit price, must also be included as birth-death terms in the Fokker-Planck equation. All quantities involved in the equation can be calibrated 
using high-frequency data on best quotes. One of our central finding is the the dynamical process is {\it approximately scale invariant}, i.e., the only relevant variable is the ratio of the
current volume in the queue to its average value. While the latter shows intraday seasonalities and strong variability across stocks and time periods, the dynamics of 
the rescaled volumes is universal. In terms of rescaled volumes, we found that the drift has a complex two-dimensional structure, which is a sum of a gradient contribution and 
a rotational contribution, both stable across stocks and time. 
This drift term is entirely responsible for the dynamical correlations between the ask queue and the bid queue. 
\end{abstract}

\smallskip



\section{Introduction}
\label{intro}

Executing orders on modern electronic, double auction markets can be achieved by posting either market orders or limit orders.\footnote{There are actually
many variation on that theme, with for example, ``Immediate or Cancel'' orders, ``Fill or Kill'' orders, ``Iceberg'' orders, etc. We are not concerned with these
subtleties in the present context.} Both market and limit orders have flaws and merits. 
Market orders are executed immediately against the best prevailing quote, but pay the half spread. Limit orders are stored in the order book 
and are only executed when a market order crosses the spread. They appear to save the half spread but face selection bias. 

It is customary to distinguish small tick and large tick situations. The tick size is the minimum amount by which the price can change. Large ticks  
correspond to stocks for which the bid-ask spread is most of the time equal to its minimum value (one tick). These relatively 
large spreads attract limit orders, naturally leading to relatively large volumes at the best quotes. 

For large tick stocks, the volume of impinging market orders is typically much smaller than the available volume at the best price. In this 
case, the rule for executing outstanding limit orders depends on the market. For most markets (including those studied in the present paper), time priority applies: 
the queue is ``first in-first out''. 
For some markets, however, a proportionality rule applies: all participants at the best quote get filled in proportion of their volume. 
If time priority applies, the best situation for a limit order is to be first in a large queue: in this case, the probability to be executed is large, 
while the adverse selection bias is small, because the large quantity of orders in the queue makes an adverse price move highly improbable. Skouras \& Farmer \cite{Skouras} 
have recently shown that if it was possible to jump the queue, huge profits would follow. Of course, this is not possible, and simply joining early is not 
enough because being alone in a queue in fact increases the probability of adverse selection. Devising efficient strategies that enable one to be well placed in large queues 
is clearly a goal pursued by all traders, be they executing brokers, market makers or high frequency traders.

The dynamics of queues in electronic markets has therefore attracted a considerable attention in the last few years. Several models have been proposed 
and analyzed in the recent literature. Many of these models are based on the simplifying assumption that the flow of market orders, limit orders and cancellations
are Poisson processes. This assumption is however clearly unwarranted for several reasons: not only because the inter-arrival times show the usual, highly intermittent patterns typical 
of the trading activity, but also because the flow of market, limit orders and cancellations are strongly intertwined \cite{Eisler}. Limit orders respond to the flow market orders
(a process called stimulated refill in \cite{molasses} or dynamical liquidity in \cite{Weber}) and vice-versa, in a way that reflects the perpetual hide-and-seek game played by 
buyers and sellers in financial markets \cite{Bence}. In order to account for some of these effects, one can use for example multidimensional Hawkes processes \cite{BacryMuzy}, equilibrium theory \cite{Rosu} 
or mean-field games \cite{CAL}. 

Another route has recently been suggested in \cite{Avellaneda,Larrard}: for large queues, one expects the changes in volume to be relatively small from one event to the next. 
Therefore, the dynamics of the queues should be captured, on a sufficiently coarse time scale, by a drift-diffusion process, parameterized by a relatively small number of 
quantities that can be calibrated on high frequency data. Rephrased in the Fokker-Planck language proposed below, the model put forth 
by Cont \& Larrard  \cite{Larrard} can 
be formulated as a {\it two-dimensional Fokker-Planck} (or diffusion) equation for the joint probability 
$P(V_A,V_B;t)$ to find a volume $V_A$ at the ask and $V_B$ at the bid at time $t$.\footnote{This equation further assumes that the bid price and the ask price have not changed 
between $0$ and $t$. See below for the inclusion of price changes in this formalism.} Their equation reads (albeit in very different notations!):
\be\label{FP2d}
\frac{\partial P}{\partial t} = - \frac{\partial (F_A P)}{\partial V_A} - \frac{\partial (F_B P)}{\partial V_B} +  
\frac{\partial^2 (D_A P)}{\partial V_A^2} + \frac{\partial^2 (D_B P)}{\partial V_B^2} + 
2 \frac{\partial^2 }{\partial V_A \partial V_B} (\rho_{AB} P),
\ee
where $P$ is a shorthand notation for $P(V_A,V_B;t)$, $F_A=F_B$ is a constant (independent of $V_A,V_B$) 
that represents the systematic drift in the evolution of the volume of the queues, while $D_A=D_B$ represent the diffusion 
coefficients in volume space, related to the variance of volume changes per unit time, chosen again to be independent of both $V_A$ and $V_B$. 
Finally, $\rho_{AB}$ is the covariance of the volumes changes on both sides of the quotes. 
It is the only term that couples the evolution of the volume at the bid and at the ask in this model. For $\rho_{AB} = 0$, the dynamics 
decouples in the sense that $P(V_A,V_B;t)$ factorizes into $P_A(V_A;t) \times P_B(V_B;t)$, with $P_A, P_B$ each obeying a one-dimensional Fokker-Planck equation:

\be\label{FP1d}
\frac{\partial P_z}{\partial t} = - \frac{\partial (F_z P_z)}{\partial V_z} + \frac{\partial^2 (D_z P_z)}{\partial V_z^2}, \qquad z=A,B.
\ee

The aim of the present paper is to include the possible dependence of the drift ($F_A$, $F_B$) and diffusion coefficients ($D_A$, $D_B$) on the volume of the two queues, $V_A, V_B$.
This dependence, neglected in \cite{Larrard}, is expected on intuitive grounds and induces correlations in the dynamics of the two queues, even when $\rho_{AB} = 0$. It is indeed 
reasonable to think that queues tend to grow when they are small and shrink when they are large, meaning that $F_z(V_z)$, $z=A,B$, is positive for small $V_z$ and negative 
for large $V_z$. One also expects that high volumes at the ask have a detrimental impact on the liquidity of the bid, and vice-versa, leading to a rich structure for the drift
field $F_{A,B}(V_A,V_B)$. These effects are indeed very clearly 
revealed by empirical data and affect quite considerably the analysis of Cont \& Larrard \cite{Larrard}, in particular concerning the calculation of the time needed to empty a queue and 
induce a price change. 

\section{Aim of the paper and main results}
\label{aim}

The present study is mostly empirical, and aims at establishing the correct model for describing the coarse-grained dynamics of the bid and ask queues, in the restricted but simpler case
where the tick is so large that the relative change of volume induced by each individual order is small. 
Our data set covers {\it all events} at the best quotes of several large tick NASDAQ stocks during the year 2010 -- see the order book animation available at http://cfm.fr/msft.exe.

We will first present (in section \ref{sec1d}) our results for the dynamics of a single queue (the bid or the ask), independently of the state of the opposite queue. 
This will allow us to present 
in a simplified setting some of our most salient results. We will investigate in particular the dependence of the drift $F_z$ and of the diffusion constant $D_z$ as a function of 
the size of the corresponding queue $V_z$. Some of our results are remarkably universal: although the trading rate and the average volume in the queues show strong intraday
seasonalities and vary significantly between different stocks, we find that upon appropriate rescalings of time and volumes, 
the statistical description of the queue dynamics is {\it independent} of the side of the queue (bid or ask), time of day, period, and considered stocks, 
provided the size of the queues is large enough (which usually entails large ticks). We find in particular 
that, as expected on intuitive grounds, $F_z(V_z)$ is positive for small $V_z$ and becomes negative for larger $V_z$. 

It turns out that the full Fokker-Planck description of single queues must involve several additional quantities. 
One describes the probability that a gap appears, momentarily increasing the bid-ask spread by one tick. Suppose one focuses on the bid, $z=B$. 
There is a chance that the next limit order that fills the newly created gap is a buy 
limit order. In that case, the new bid jumps one tick up. This happens with a certain probability $Q_+$ that may again depend on $V_B$, and when this happens, the new bid
starts with a volume distributed according to a certain $P_+(V_B)$. Conversely, the volume at bid can be eaten by a large sell market order, in which case the bid goes down 
one tick, and is replaced by the queue just behind it. This happens with probability $Q_-(V_B)$ and when this happens, the new bid starts with a volume distributed according 
to a certain $P_-(V_B)$. Similar quantities describe the same events at the ask, $z=A$. Our convention will be to use a $+$ subscript for events that {\it improve} the quotes 
(bid up or ask down one tick), and a $-$ subscript for events that {\it degrade} the quotes (bid down or ask up one tick). Again rescaling times and volumes, we find that these
new quantities are again to a large degree universal. Finally, we test directly the validity of our Fokker-Planck description by comparing the empirically determined 
stationary distribution of volumes $P_{st}(V_A)$ (or $P_{st}(V_B)$) with the one predicted by the equilibrium solution of the Fokker-Planck equation.

We then turn in section \ref{sec2d} to the full two-dimensional description of the dynamics. We find (empirically) a complete statistical symmetry between the bid and the ask, in particular that 
the stationary distribution obeys $P_{st}(V_A,V_B)=P_{st}(V_B,V_A)$, while the drifts and diffusion coefficients are such that $F_A(V_A,V_B)=F_B(V_B,V_A)$ and 
$D_A(V_A,V_B)=D_B(V_B,V_A)$, and similarly for $Q_\pm$ and $P_\pm$, that describe change of prices. We again find that upon adequate rescaling, all these quantities are universal. 
In view of the rather subtle pattern revealed by the drift field $(F_A,F_B)$ (see Figs. 8 \& 9 below), this universality is far from trivial.

As pointed out by Cont \& Larrard \cite{Larrard}, a model for the dynamics of queues is valuable for many purposes. One of the interesting quantity that our model allows one to compute (in principle) is 
the probability that, starting from a volume configuration $(V_B,V_A)$, the queue bid moves down before the ask bid moves up, or the distribution of times before the bid price moves, etc.
Due to the complexity of the calibrated model, this can however only be achieved by running numerical simulations, and we leave this question for future investigations (see also the discussion in
Sect. 6 below).

\section{Data \& descriptive statistics}
\label{data}

Our data consists in all events on the NASDAQ platform (market orders, limit orders, cancellations) occurring at the best quotes (bid/ask) for NASDAQ stocks during the year 2010. 
These orders represent only $\approx 40 \%$ of the total activity, but in view of the universality of our results once rescaled by the appropriate average volume, 
we believe that our conclusions are not significantly affected by the missing 
data. We do not consider iceberg limit orders either, which is consistent with our assumption that the queue dynamics is only affected by the visible volume on both queues.
For reasons explained above, we only focus on large tick stocks. Since the value of the tick on the NASDAQ is constant and equal to 0.01 USD, large tick stocks means stocks with
relatively small prices per share. Our criterion was to choose stocks for which the average price in 2010 was $ < 40$ USD, i.e. ticks $> 2.5$ basis points (see Table 1). 
Although we checked that our results are qualitatively valid for all these large tick stocks listed in the table, we have only done an extensive analysis of the data for five of these stocks: 
Microsoft, Oracle, Gilead Sciences, Cisco Systems and eBay -- see Table I.

\begin{table}
\begin{tabular}{|c|c||c|c|c|c|c|c|c|}
\hline
Stock Names             & Ticker & Av. Price & $\bar{L}$ & $\bar{V}$ & $\bar{N}$ & $\langle|V|\rangle$ & $\bar{\Pi_0}$ &  $\pi_+$\\ \hline
Cisco Systems, Inc.     & CSCO   & 23.2  & 44.2      & 25,000 & 2,240 & 900 & 0.86 & 0.19  \\ \hline
eBay                    & EBAY   & 24.6  & 17.2      & 4,900 & 1,240  & 475  & 0.87 & 0.15\\ \hline
Gilead Sciences, Inc.   & GILD   & 39.3  & 11.1      & 2,350 & 1,140  & 335  &  0.85 & 0.21 \\ \hline
Microsoft Corporation   & MSFT   & 27.1  & 44.8      & 22,100 & 2,630 & 890 & 0.90 & 0.21\\ \hline
Oracle Corporation      & ORACL  & 24.7  & 29.6      & 12,800 & 1,800 & 730 & 0.87 & 0.16 \\ \hline
\end{tabular}
\caption{Summary statistics for the 5 stocks chosen for this study. The average price (in USD) is over the year 2010. $\bar{L}$ (resp. $\bar{V}$) corresponds to the 
average number of individual orders (resp. average volume in shares) in the queue at any instant of time (see Fig. 1 for the intraday pattern). $\bar{N}$ is the average 
number of events that modify $L$ and $V$ during a a 5 minutes interval.  $\langle|V|\rangle$ is the average (absolute) change of volume for each event that does not 
change the price. $\bar{\Pi_0}$ is the average probability that an event does {\it not} change the best level 
(see Fig. 3 for the intraday pattern). $\pi_+$ is the average probability that a freshly emptied queue is immediately refilled, meaning that the best quote does not change after
such an event. Note that $\pi_+$ is around $0.20$ for most of the day, but with sharp peaks around the open and the close, when $\pi_+$ reaches $0.35 - 0.40$.}
\end{table}

We will denote by $L$ the size of the queue measured in number of different individual orders (which can all be of different volumes), $V$ the size of the queue in total volume (i.e. 
number of shares),  and $N$ the number of events that modify $L$ and $V$ during a specific period, in our case a 5 minutes bin. The trading day is therefore divided into 78 bins of 5 minutes. 
$L$ and $V$
give slightly different informations about the size of the queue, and the Fokker-Planck formalism could be applied for each of these two variables. We have in fact 
studied both cases, with very similar conclusions \cite{Gareche_Report}. However, taking the total {\it volume} $V$ leads to less noisy, more regular observables and probably 
makes more financial sense, so we restrict in this paper to volumes only and will write an evolution equation for $P(V_A,V_B;t)$, the joint probability to observe volumes $V_A$ 
at the ask and $V_B$ at the bid at ``time'' $t$, where time will be counted here in {\it event time}.

The first interesting information is to characterize the average daily pattern of the activity $N$ and size of the queues $L_{A,B}$ or $V_{A,B}$. Averaging over all days and 
all stocks, we obtain the characteristic patterns shown in Fig. 1 for $\bar{L}(b)$ and $\bar{V}(b)$, where $b=1,2,\dots,78$ is the bin number. For the total activity $\bar{N}(b)$, 
we find the familiar $U$-shape (not shown) : activity is high in the morning, lower at noon, and high again at the end of the day. The total volume in the book, on the other hand, 
is quite low at the open and steadily rises as one moves into the day, with an interesting acceleration towards the end of the day. The plots shown in Fig. 1 are averages over the
bid and the ask, {\it and} averages over the 5 chosen stocks, but we have checked that the individual patterns for $\bar{L}(b)$, $\bar{V}(b)$ and $\bar{N}(b)$
are the same up to an overall multiplicative factor. 
While the rescaling is not perfect and some differences of order $\sim 20 \%$  between stocks might be 
relevant for a finer analysis, we are content with the idea that in a first pass, universality holds.\footnote{The difference between the bid and ask observables is of the same 
order of magnitude, but it is highly reasonable that the statistics of high frequency activity should be very close to being buy/sell symmetric.} 

\begin{figure}
\begin{center}
 \includegraphics[scale=0.3]{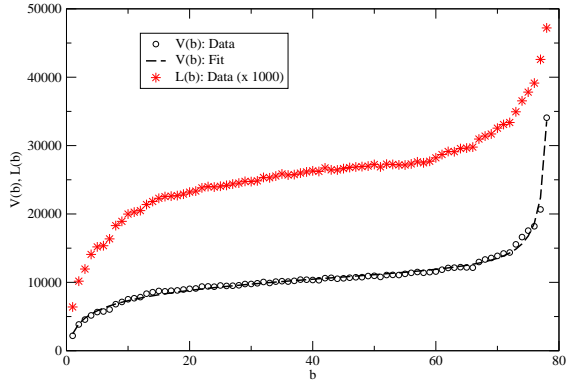}
  \caption{Intraday pattern of the average volume $\bar{V}(b)$, and average number of orders $\bar{L}(b)$ in the queue, and fit with Eq. (3). Averages are over all days and
  all five stocks.
  Note that both quantities are in fact close to being proportional to 
  each other, with an average (over all stocks) volume per order $\approx 400$.}
\end{center}
 \end{figure}

Interestingly, the intraday pattern of the volume is accurately fitted by the following simple functional form (see Fig. 1):
\be
\bar{V}(b) \approx a_0 + a_1 \ln b + \frac{a_2}{79-b}.
\ee
The meaning of this functional form is that the volume increases quickly at the beginning of the day before a quasi-plateau, as captured by the initial
logarithmic dependence, before the final increase that shows an apparent divergence at the end of the day. We note that
such an inverse divergence of the activity close to a deadline (here the end of the day) has been reported for other human activities as well \cite{pietronero}. It would be
quite interesting to adapt the behavioral pattern of \cite{pietronero} in the present context, maybe related to market-makers liquidating their position before market close.\footnote{Note that leaving the exponent $\psi$ of the divergence 
as a free parameter, a best fit of the data leads to $\psi \approx 1.05$.}

Let us now turn to what will be the central object of the present study, namely the probability $P_{st}(V_A)$ that the ask queue has volume $V_A$ (or similarly for the bid volume $V_B$).
Clearly, since the average volume $\bar{V}$ depends on the time of the day and on the stock, $P_{st}$ cannot be universal. Our central assumption, that is approximately borne out by the 
data, is that for large queues, all aspects of the queue dynamics {\it only depends on relative volumes}, i.e. on the ratio of the existing volume $V_A$ over the average volume $\bar{V}$.
In other words, introducing $x_A=V_A/\bar{V}$ and $x_B=V_B/\bar{V}$, one expects that $P_{st}(x)$ is approximately universal, both in time and across stocks. As we will see later, this 
assumption naturally generalizes to other quantities as well. 

We show $P_{st}(x)$ in Fig. 2-a for the five stocks under scrutiny (averaged over the bid and at the ask). 
For all stocks, we observe a hump shaped distribution that peaks around the average value $\bar{x}=1$. The probability of much larger queues goes to zero slightly slower than 
exponentially. Of course, the joint 
distribution of $x_A=x$ and $x_B=y$ contains more information, and is shown as a contour plot in Fig. 2-b, here averaged over all 5 stocks. We find, quite interestingly, that 
$P_{st}(x,y)$ exhibits a broad peak around $x \approx y \approx 1$. This means that the most probable situation is that both queues are of similar height, with a 
an average volume $\bar{V}(b)$ that is bin- and stock- dependent. (see also the animation available at http://cfm.fr/msft.exe).

\begin{figure}
\begin{center}
 \includegraphics[scale=0.24]{IndividualP.eps}\hskip 1cm \includegraphics[scale=0.28]{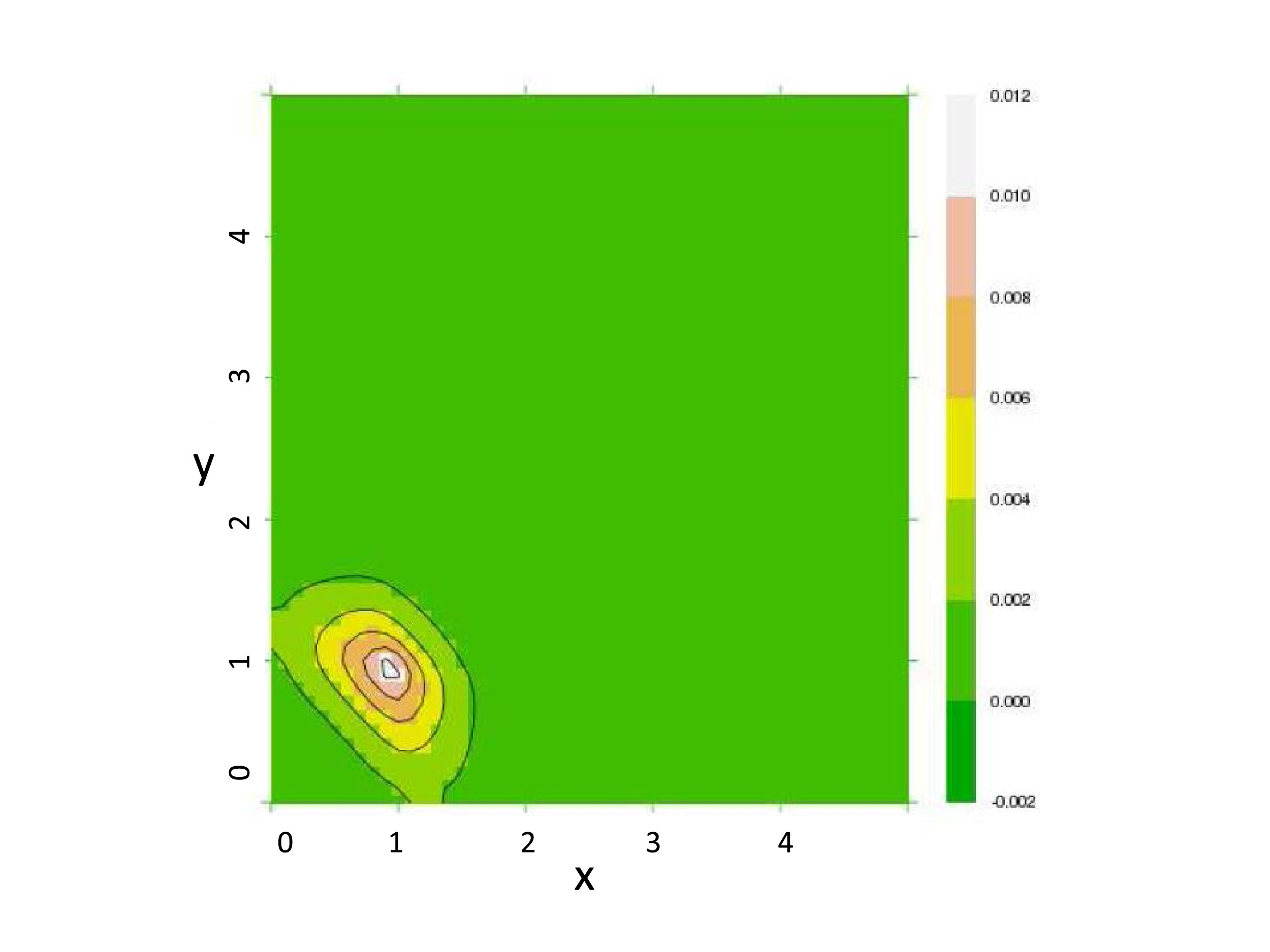} 
  \caption{Left: Individual $P_{st}(x)$ for the 5 stocks studied, obtained by averaging over all days the distribution of the rescaled variable $V/\bar{V}(b)$.
  The $y$-axis is in log scale.
  Right: Two dimensional joint distribution of rescaled volumes at the bid and the ask, shown as contour levels of $P_{st}(x,y)$, which exhibits a broad peak around $x \approx y \approx 1$. 
  Note the symmetry around the line $x=y$.
  }
\end{center}
 \end{figure}

What are the dynamical mechanisms at the origin of these specific, humped shaped distributions? The aim of the following sections is to develop a precise picture of the stochastic
process governing the joint evolution of the two queues. We first focus on a one-dimensional model, that discards all information about the opposite queue, before expanding on the 
full two dimensional model in the following section.

\section{A one dimensional model}
\label{sec1d}

\subsection{Derivation of the Fokker-Planck equation}

We start by writing a general Master equation for the evolution of the probability that the volume at the bid or at the ask is equal to $V$, at (event) time $t$. Assuming for the moment that there is no change of the corresponding price between $t$ and $t+1$, this reads:
\be
P(V,t+1) = \sum_{\Delta V} P(V-\Delta V,t) \rho(\Delta V|V-\Delta V),
\ee
where $\Delta V$ are the possible changes of volume related to limit orders ($\Delta V > 0$) and market orders or cancellations ($\Delta V < 0$), that occur with a $V$ dependent 
probability $\rho(\Delta V|V)$. The above Master equation assumes that the process is Markovian, i.e. no memory in time, apart from the one encoded in the instantaneous size of the queue. When $V$ is large, one may expect that 
changes of the queue size at each time step is relatively small: $\Delta V \ll V$. This allows one to treat $V$ as a continuous variable,  expand the above equation in powers of $\Delta V$. 
The general expansion of the Master equation in powers of $\Delta V$ is called the Kramers-Moyal expansion; when truncated to second order, this leads to the Fokker-Planck equation \cite{Gardiner}. 
In the present case, one finally gets:
\be\label{FP-1}
P(V,t+1)-P(V,t) \approx  - \frac{\partial [F(V) P(V,t)]}{\partial V} + \frac{\partial^2 [D(V) P(V,t)]}{\partial V^2};
\ee
with:
\be
F(V) = \sum_{\Delta V} \Delta V \rho(\Delta V|V); \qquad D(V) = \frac12  \sum_{\Delta V} (\Delta V)^2 \rho(\Delta V|V);
\ee
in other words, $F(V)$ is the average volume change conditional to a certain volume $V$, whereas $D(V)$ is the (one-half) of the average volume change squared, again
conditioned to $V$. 

There are two additional processes that need to be taken into account in order to faithfully describe the dynamics of queues -- say the bid.
\begin{itemize}
\item One is that the opposite ask moves up one tick, which leads to a situation where the spread between the bid and the ask is temporarily equal to two ticks. 
If a new buy limit order fills the incipient gap the `old' bid, with volume $V$ then gets suddenly replaced by a new bid, 
with (usually) smaller volume. In the Master equation language, this corresponds 
to a large jump for which the assumption that $\Delta V$ is small is not warranted. 
We instead want to model this effect by adding to the right hand side of Eq. \ref{FP-1} a ``birth-death'' term of the form:
\be\label{jump}
- Q_+(V) P(V,t) + \left[\sum_{V'} Q_+(V') P(V',t)\right] P_+(V),
\ee
where $Q_+(V)$ is the probability that a queue of size $V$ gets overtaken by a new queue at an improved price, 
and $P_+(V)$ is that probability that a newly created queue starts with volume $V$.\footnote{Note that $Q_+(V)$ does {\it not} count events where the opposite quote 
disappears and immediately reappears at the same price, leaving the considered quote unchanged. In other words, $Q_+(V)$ already includes the probability $\pi_+$.}

\item The second effect is that when the bid has a small volume, there is a finite probability that the bid is eaten entirely by a market order or by a cancellation. 
Two things can happen in this case: either the queue one tick below the old bid becomes the new bid, or some volume immediately comes back with no price change.\footnote{Formally,
this corresponds to two events. However, we find it more consistent to restrict the state space of the model to situations where the spread is equal to one tick, and remove from
the description the highly transient situations where the spread is equal to two ticks.} Both cases again correspond to ``jumps'' in the Fokker-Planck framework. 
We write that with probability $Q_-(V)$ the old queue is completely eaten. With probability $\pi_- \times P_-(V')$ it is replaced by the queue just below of size $V'$, and with probability
$\pi_+=1-\pi_-$ some new volume $V'$ reappears at the same price, with probability $P_+(V')$. The ``birth-death'' term now reads:
\be\label{jump2}
- Q_-(V) P(V,t) + \left[\sum_{V'} Q_-(V') P(V',t)\right] [\pi_+ P_+(V) + \pi_- P_-(V)],
\ee 
This is described by exactly the same term as in Eq. (\ref{jump}) above, with $Q_+ \to Q_-$ and $P_+ \to \pi_+ P_+(V) + \pi_- P_-(V)$, with $\pi_+ \approx 0.15 - 0.2$, see Table I.
\end{itemize} 

Note however that these price changing events impose that the distribution of volume changes, $\rho(\Delta V|V)$ is not normalized to unity rather but to 
$\Pi_0(V)=1 - Q_+(V) - Q_-(V)$, the probability that an event does not change the price.
In the following, we will use the notation $F(V),D(V)$ for the average drift and diffusion conditional to no price change, and $\tilde F(V),\tilde D(V)$ for the unconditional
quantities, with:
\be
\tilde F(V) = \Pi_0(V) F(V); \qquad \tilde D(V) = \Pi_0(V) D(V).
\ee
Empirically, $\Pi_0(V)$ is in fact found to be $\approx 0.9$, see Fig. 3 and Table 1.

\begin{figure}
\begin{center}
 \includegraphics[scale=0.3]{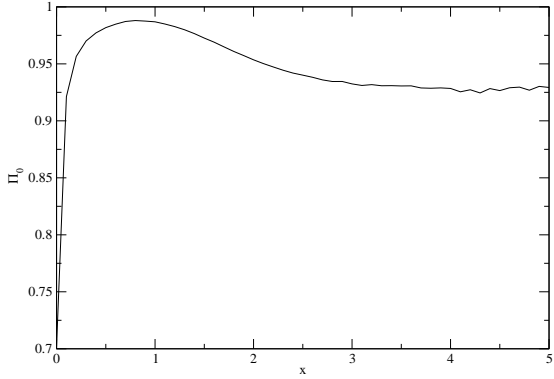}
  \caption{Dependence of the probability of non price-changing events $\Pi_0(x)$ on the rescaled variable $x=V/\bar{V}$, and averaged over all 5 five stocks.
  Note the dip for small volumes in the queue, which have a large probability to be eaten by a single market order.}
\end{center}
 \end{figure}

\subsection{Normalization}

As discussed above, it is reasonable to expect that the volume dynamics is, for large queues, scale-invariant, in the sense that only the ratio $x=V/\bar{V}$ matters, where
$\bar{V}$ is the average volume in the queue, which is both stock- and time-of-day-dependent. It is easy to see that the Fokker-Planck equation, in terms of the rescaled volume $x$, takes 
the following form:
\bea\label{FP-final}
P(x,t+1)-P(x,t) \approx  - \frac{\partial [\tilde f(x) P(x,t)]}{\partial x} + \frac{\partial^2 [\tilde d(x) P(x,t)]}{\partial x^2}- q_+(x) P(x,t) +\\
+ \left[\sum_{x'} q_+(x') P(x',t)\right] P_+(x)- q_-(x) P(x,t) + \left[\sum_{x'} q_-(x') P(x',t)\right] [\pi_+ P_+(x) + \pi_- P_-(x)]
\eea
with:
\be
\tilde f(x):=\frac{\tilde F(x\bar{V})}{\bar{V}}-\frac{x}{\bar{V}}\frac{d\bar{V}}{dt},
\ee
\be
\tilde d(x)=\frac{\tilde D(x\bar{V})}{\bar{V}^2}; \qquad q_\pm(x):=Q_\pm(x \bar{V}),
\ee
and all probability densities such that, in a shorthand notation, $P(x)dx = P(V)dV$. 
Note that Eq. (\ref{FP-final}) is such that $\sum_{x} P(x,t+1) = \sum_{x} P(x,t)$, as it should be.

Eq. (\ref{FP-final}) is the central equation of this work, and defines in a precise manner our model for single queue dynamics. All the information needed to determine the input of this 
equation (namely the functions $f(x),d(x),q_\pm(x)$ and $P_\pm(x)$), can be precisely calibrated on data. Indeed, thanks to the simplifying {\it scale-invariance assumption}, 
all these quantities can be determined by aggregating data at different times of the day for a single stock, and by further averaging over different stocks. Again, there might be slight 
inter-stock variations, or some dependence on the specific period of time, but our detailed analysis of the data has convinced us that {\it as a first approximation}, the scale-invariance
property holds with reasonable accuracy. More work is needed to ascertain whether these variations are statistically significant, but this is well beyond the scope of the present study.

\subsection{Empirical analysis}

Based on the above scaling assumption, we have determined the functions $f(x),d(x),q_\pm(x)$ and $P_\pm(x)$ on our data set, averaging over all 5 stocks mentioned above, all days of 2010
and all 78 bins of each day. The results are shown in Fig. 4 and Fig. 5. Fig. 4 reveals intuitive, but interesting results: we find that, as expected, the drift $f(x)$ is negative 
for large $x$'s, meaning that long queues (as compared to the average value) are shrinking, whereas short queues are expanding. In fact, we find that the drift vanishes when $x \approx 1$, 
i.e. for average-sized queues. The coefficient $d(x)$ essentially measures the intensity of activity in the queue; it is found to decrease slightly between $x=0$ and $x=1$, before gradually
increasing and becoming $10$ times larger for $x \approx 4$. The quantities $q_\pm(x)$ and $P_\pm(x)$ are plotted in Figs 5-a and 5-b, respectively. One observes that $q_-(x)$ 
reaches a minimum for typical queue sizes ($x \sim 1$), meaning that it is quite rare that these queues get eaten by a single trade. This probability is much higher for small queues 
and for large queues. In the former case, this comes from the fact that traders try to grab small volumes at the best before the price moves adversely. In the latter case, large queues 
offer opportunities to execute large orders in a single shot, thereby limiting impact costs. The probability that a new queue improves the current best, $q_+(x)$ is seen to increase 
monotonically as a function of the size of the queue. This is quite expected: if a gap opens in front of the current best, the incentive to place an order there rather than to
join the queue increases as its volume $x$ increases. The plots of $P_\pm(x)$ shown in Fig. 5-b are also not surprising: the distribution of the volume at the 
second best level ($P_-(x)$) is similar to the unconditional distribution of the best ($P(x)$), whereas the distribution of volume at incipient levels ($P_+(x)$) is strongly peaked at $x=0$. 

 \begin{figure}
\begin{center}
 \includegraphics[scale=0.3]{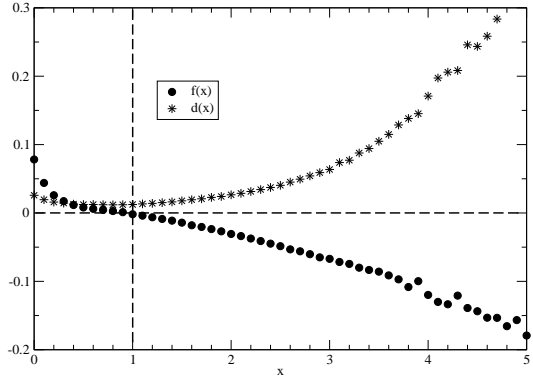}
  \caption{Drift $f(x)$ and diffusion $d(x)$, conditioned to no price change, as a function of the rescaled volume $x$. Note that, as indicated by the dotted horizontal and vertical lines, $f(x=1) \approx 0$. For large $x$, $d(x)$ 
  increases by a factor $10$ or more compared to the value $d(x=1)$.}
\end{center}
 \end{figure}
 
  \begin{figure}
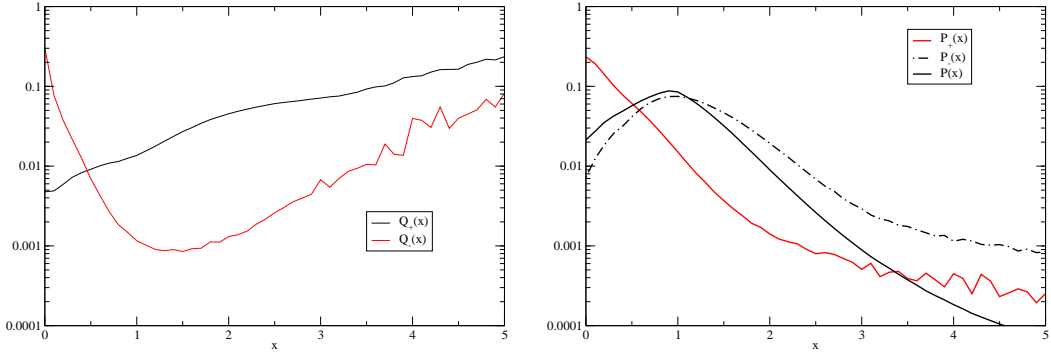

 \begin{center}
  \includegraphics[scale=0.28]{Q+Q-.eps}\hskip 0.5cm \includegraphics[scale=0.28]{P+P-.eps}
   \caption{Left: Probability that the current queue disappears by being overtaken ($q_+(x)$) or by being completely eaten ($q_-(x)$), both as a function of the
   rescaled volume $x$. Right: Probability that the newly appeared volume is $x$, at a better price ($P_+(x)$) or at a worse price ($P_-(x)$).}
 \end{center}
  \end{figure}

\subsection{Discussion}

Our final model, Eq. (\ref{FP-final}), is based on two major assumptions. One is that the dynamics is Markovian, i.e. it  depends on the past flow of order {\it only through the current 
size of the queue} (relative to its average for a given stock and a given hour of the day). In other words, one assumes that there is no temporal correlation in the type and volume of events.
The second assumption is that one can assume the change of volume to be small for all events that do not change the price of the bid/ask. 

\begin{figure}
\begin{center}
 \includegraphics[scale=0.25]{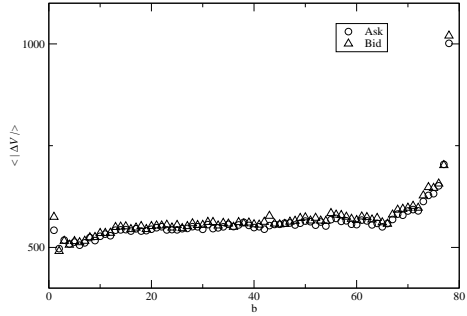}
  \caption{Dependence of $\langle | \Delta V | \rangle$ as a function of the time of day, in 5-minute bin units, averaged over all stocks (see also Table I).}
\end{center}
 \end{figure}

We have checked that the event correlations is small, so the first assumption appears to be warranted. The second assumption is however not completely justified: as shown in Fig. 6  the typical volume change $\Delta V$ is $\sim 500$, which is 30 times smaller than the average volume $\bar{V} \sim 15,000$ (see Table I for individual stock statistics). However, the distribution of $\Delta V$ has heavy (power-law) tails, which means that higher order derivatives in the Kramers-Moyal expansion could play a role and invalidate the Fokker-Planck truncation.  

We propose to check directly of the validity of the Fokker-Planck approximation by using Eq. (\ref{FP-final}) with the empirically determined inputs ($f(x),d(x),q_\pm(x)$ and $P_\pm(x)$) to reconstruct the stationary distribution $P_{st}(x)$. The steady-state equation reads:
\bea\label{FP-st}
- \frac{\partial [\tilde f(x) P_{st}(x)]}{\partial x} + \frac{\partial^2 [\tilde d(x) P_{st}(x)]}{\partial x^2} = q_+(x) P_{st}(x) - \left[\sum_{x'} q_+(x') P_{st}(x')\right] P_+(x)\\
 + q_-(x) P_{st}(x) - \left[\sum_{x'} q_-(x') P_{st}(x')\right] [\pi_+ P_+(x) + \pi_- P_-(x)].
\eea
There is no general analytic solution for $P_{st}(x)$. However, when $q_\pm(x)=0$, the zero-current solution is simply given by the Gibbs-Boltzmann measure:
\be\label{GB}
P_{GB}(x) \propto \frac{1}{d(x)} \exp\left[-u(x)\right]; \qquad u(x) = -\int_0^{x} {\rm d}x' \frac{f(x')}{d(x')},
\ee
where $u(x)$ is the ``potential'' and $d(x)$ can be interpreted as a local, $x$-dependent temperature.  Since $1-\bar{\Pi_0} \sim 0.1$ (see Table 1 and Fig. 3, neglecting the price changing processes, i.e. setting $q_\pm(x) = 0$ should be a reasonable approximation. The result is shown in Fig. 7. This approximation captures well the overall humped shape of $P_{st}(x)$. This is a direct consequence of the fact that the drift $f(x)$ vanishes for $x=1$, corresponding to a minimum of $u(x)$. 

Although not perfect, the agreement between the empirical distribution $P_{st}(x)$ and the 
reconstructed Gibbs-Boltzmann measure $P_{GB}(x)$ is far from trivial, since the quantities $f(x)$ and $d(x)$ needed to reconstruct $P_{GB}(x)$ are measured from the {\it dynamics} of the queue.
We believe that this approximate agreement, with {\it no extra fitting factor}, is a convincing empirical validation of our Fokker-Planck formalism.

\begin{figure}
\begin{center} 
 \includegraphics[scale=0.28]{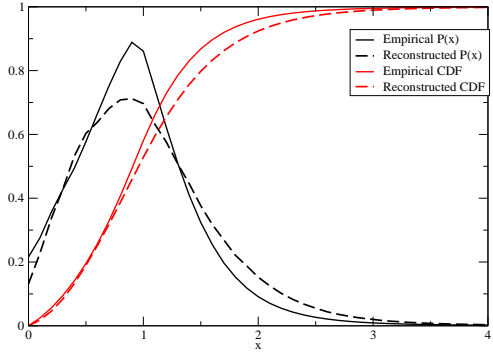} 
  \caption{Empirical $P_{st}(x)$, averaged over the 5 stocks, and reconstructed $P_{GB}(x) \propto d^{-1}(x)\, e^{-u(x)}$. The $y$-axis is in log scale. We have represented 
  for clarity both the pdfs $P(x)$ and the cumulative distribution functions $\int_0^x {\rm d}x' P(x')$.}
\end{center}
 \end{figure}

Finally, we note that the Fokker-Planck equation in the continuous time limit is equivalent, between two price jumps, to a Brownian motion model for the rescaled queue size $x$, given by:
\be
{\rm d}x = f(x) {\rm d}t + \sqrt{2d(x)} {\rm d}W,
\ee
where ${\rm d}W$ is the standard Wiener noise. This is interesting for numerical simulation purposes. Suppose for example one starts in a situation where the bid queue is at $x=x_0$ for $t=0$ and 
ask for the probability that the queue empties while always remaining at the same price. One can integrate the diffusion equation above with initial condition $x=x_0$, adding the possibility of price changing processes at each time step. With probability $q_+(x) {\rm d}t$ the best price is improved, in which case the price goes up, and the process stops. With probability $q_-(x) {\rm d}t$, on the other hand, a large volume eats the whole queue, the price goes down and the process also stops. Finally, if the walk may survive and reach $x=0$ for the first time at $t$; this event contributes to the probability one wants to compute.

\section{The two-dimensional model}

\label{sec2d}

\subsection{A two-dimensional reduced form Fokker-Planck equation}

Generalizing the above arguments to the joint dynamics of the bid volume $V_B$ and ask volume $V_A$, we introduce relative volumes 
$x=V_B/\bar{V}$ and $y=V_A/\bar{V}$. (Note 
again that the stock and time dependent average volume $\bar{V}$ is the same for the bid and the ask). The scale-invariant, two-dimensional Fokker-Planck equation
now reads:
\bea\label{FP-2d-final}
P(x,y,t+1)-P(x,y,t) \approx  - \frac{\partial [\tilde f_x(x,y) P(x,y,t)]}{\partial x} - \frac{\partial [\tilde f_y(x,y) P(x,y,t)]}{\partial y}+
\frac{\partial^2 [\tilde d_x(x,y) P(x,y,t)]}{\partial x^2}  +\\ \nonumber
+ \frac{\partial^2 [\tilde d_y(x,y) P(x,y,t)]}{\partial y^2} - [q_+(x|y)+q_+(y|x)] P(x,y,t) + \left[\sum_{x',y'} [q_+(x'|y') P(x',y',t)\right] P_+(x,y) +\\ \nonumber
 + \left[\sum_{y',x'} [q_+(y'|x') P(x',y',t)\right] P_+(y,x) - [q_-(x|y)+q_-(y|x)] P(x,y,t) + \left[\sum_{x',y'} q_-(x'|y') P(x',y',t)\right] \times 
\\ \nonumber
\times [\pi_+ P_+(x,y) + \pi_- P_-(x,y)] +  \left[\sum_{y',x'} q_-(y'|x') P(x',y',t)\right] [\pi_+ P_+(y,x) + \pi_- P_-(y,x)],
\eea
where $\tilde f_{x,y}(x,y)$ is the average drift of the rescaled bid/ask volume, conditioned to a certain $(x,y)$, and $\tilde d_{x,y}(x,y)$ is the diffusion constant in the 
$x$/$y$ direction, again conditioned to a certain $(x,y)$. In order to be precise, we specify these definitions as follows:
\be
\tilde f_x(x,y):=\frac{\tilde F_x(x\bar{V},y\bar{V})}{\bar{V}}-\frac{x}{\bar{V}}\frac{d\bar{V}}{dt},\quad 
\tilde F_x(x\bar{V},y\bar{V})=\frac{\Pi_0}{2} \sum_{\Delta V_B} \Delta V_B \rho(\Delta V_B|V_B,V_A);
\ee
and
\be
\tilde d_x(x,y)=\frac{\tilde D(x\bar{V},y\bar{V})}{\bar{V}^2}; \quad \tilde D_x(x\bar{V},y\bar{V})=\frac{\Pi_0}{4} \sum_{\Delta V_B} (\Delta V_B)^2 \rho(\Delta V_B|V_B,V_A).
\ee
Note the extra factor $1/2$ coming from the fact that each event can occur with probability $1/2$ at the ask and $1/2$ at the bid, and note that $\rho(\Delta V_B|V_B,V_A)$ is normalized
to the probability of no price changing events for a given $V_B,V_A$. By symmetry, we expect (and have indeed confirmed empirically) that $f_x(x,y)=f_y(y,x)$ and 
$d_x(x,y)=d_y(y,x)$.

The quantities $q_\pm(x|y)$ are, respectively, the probability that, for the next event, a queue of rescaled volume $x$, facing a queue of rescaled volume $y$, 
disappears entirely ($q_-$) or gets superseded ($q_+$) by a new queue. Correspondingly, the quantity $P_-(x|y)$ gives the probability that the second best 
queue that becomes the best queue has volume $x$, knowing that the opposite queue has volume $y$, whereas $P_+(x|y)$  gives the probability that the newly 
created best has volume $x$, knowing that the opposite queue has volume $y$. (In order to simplify the presentation, Eq. (\ref{FP-2d-final}) in fact assumes, in line with our 
empirical results, a total bid/ask symmetry for the statistics of these price changing events, e,g. $q_+^B(x|y)=q_+^A(x|y)$). 

Finally, note that the mixed diffusion term $\rho \partial^2/\partial x \partial y$ originally introduced by Cont \& Larrard \cite{Larrard} is not present in the above equation. This is because we have not 
found any significant correlations between the fluctuations of volume changes at the bid and at the ask, that would justify the presence of such a term. 
However, this does {\it not} mean that we neglect the coupling between the two queues, which is entirely encoded in the two dimensional drift field $\vec f=(f_x,f_y)$, which is a function of the volumes on both sides (see Fig. 8 below). 

We now present an empirical determination of these two-dimensional quantities on the same data set as above. 

\subsection{Empirical analysis}

Based on our scaling assumption, we again determine the two dimensional functions $f_x,f_y,d_x,d_y$, $q_\pm$ and $P_\pm$ on our data set, 
averaging over all 5 stocks mentioned above, all days of 2010 and all 78 bins of each day. 
We find (see Fig 8-a) that the diffusion coefficients are, in fact, to a very good approximation {\it independent} of the size of the opposite queue, i.e.:
\be
d_x(x,y) = d(x), \quad \forall x, \qquad d_y(x,y) = d(y), \quad \forall y,
\ee
where $d(.)$ is the one-dimensional diffusion coefficient. This independence of $d_{x,y}$ on the size of the opposite queue is compatible with the absence of correlation of the 
fluctuations of activity on the bid and on the ask.

\begin{figure}
\begin{center}
 \includegraphics[scale=0.27]{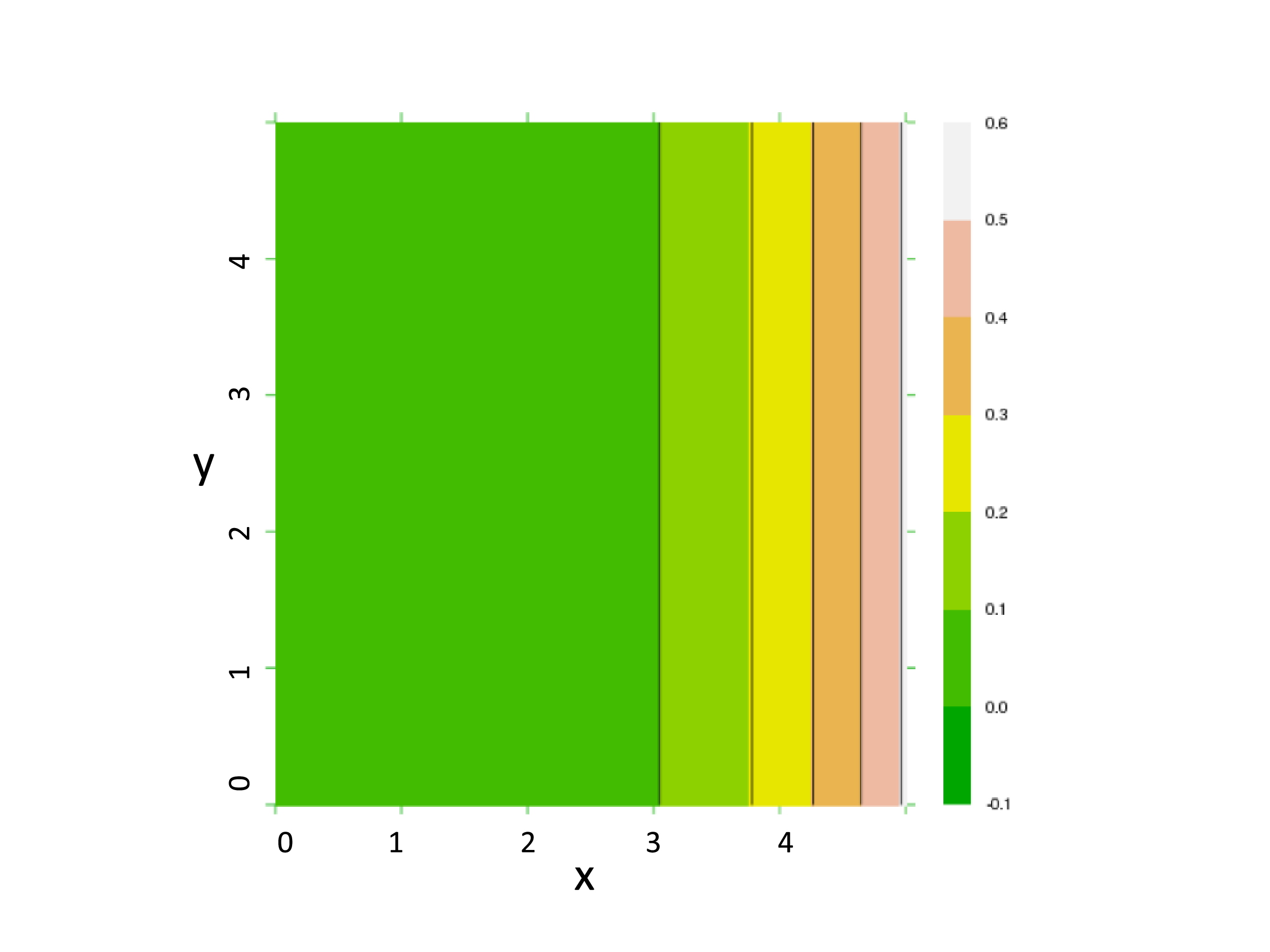}\hskip 0.5cm \includegraphics[scale=0.3]{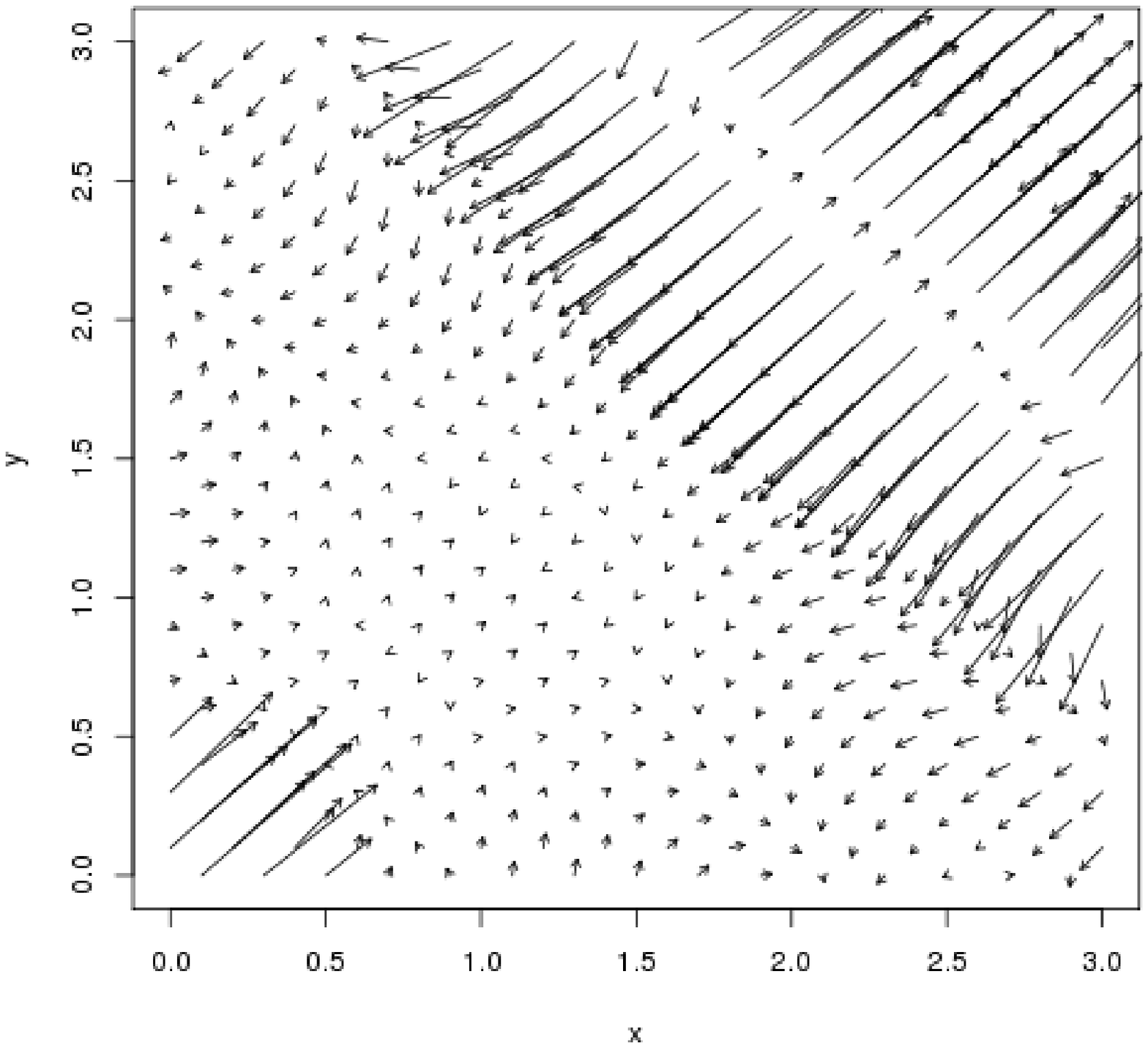}
  \caption{Left: Diffusion coefficient of the bid queue, $d_x(x,y)$, as a function of the (rescaled) bid volume $x$ and the ask volume $y$. 
  This level representation makes it clear that the diffusion coefficient is independent of the volume of the opposite queue. Right: Arrow representation of the drift field $\vec f(x,y)$ in two-dimensions. Note the bid-ask symmetry that implies $f_x(x,y)=f_y(y,x)$. While $\vec f \approx 0$ in the most 
  probable region $x \approx y \approx 1$, the drift is to a good approximation parallel to the diagonal $\vec e=(1,1)$. }
\end{center}
 \end{figure}

The structure of the drift is much more complex and interesting. For each configuration $(x,y)$ of the rescaled queue volumes, one can determine the two dimensional drift 
vector $\vec f=(f_x,f_y)$, which is represented as arrows in Fig. 8-b. This is obtained as a grand average across time and across stocks, but we have found that the pattern reported
in Fig. 8-b is actually the same for different stocks, or when we divide the 2010 time period in monthly sub-intervals, or else when we focus on morning hours or afternoon hours \cite{Gareche_Report}. 
What makes this universality possible at all is of course that we work with rescaled volumes. (Very similar patterns appear if one works with $L$, the number of different orders in the
queue, rather than $V$, the total volume.) One sees a pattern recalling, at first glance, the one-dimensional situation: large queues tend to shrink while small queues tend to grow, with 
a central region around $x \approx 1, y \approx 1$ where the drift is small (and thus noisy) -- see the animation available at http://cfm.fr/msft.exe. A better way to visualise the drift is through the introduction of potentials. In two dimensions, a 
vector field can be uniquely decomposed as the sum of a potential field and a rotational field, i.e.:
\be
\vec f = - \vec \nabla u + \vec \nabla \times \vec w,
\ee
where $u(x,y)$ is the potential (similar to the one-dimensional object above) and $\vec w=(0,0,w)$ is a vector orthogonal to the $x,y$ plane. This second, rotational part, contributes to closed current loops in equilibrium, whereas the potential part does not. 

 \begin{figure}
 \begin{center}
  \includegraphics[scale=0.30]{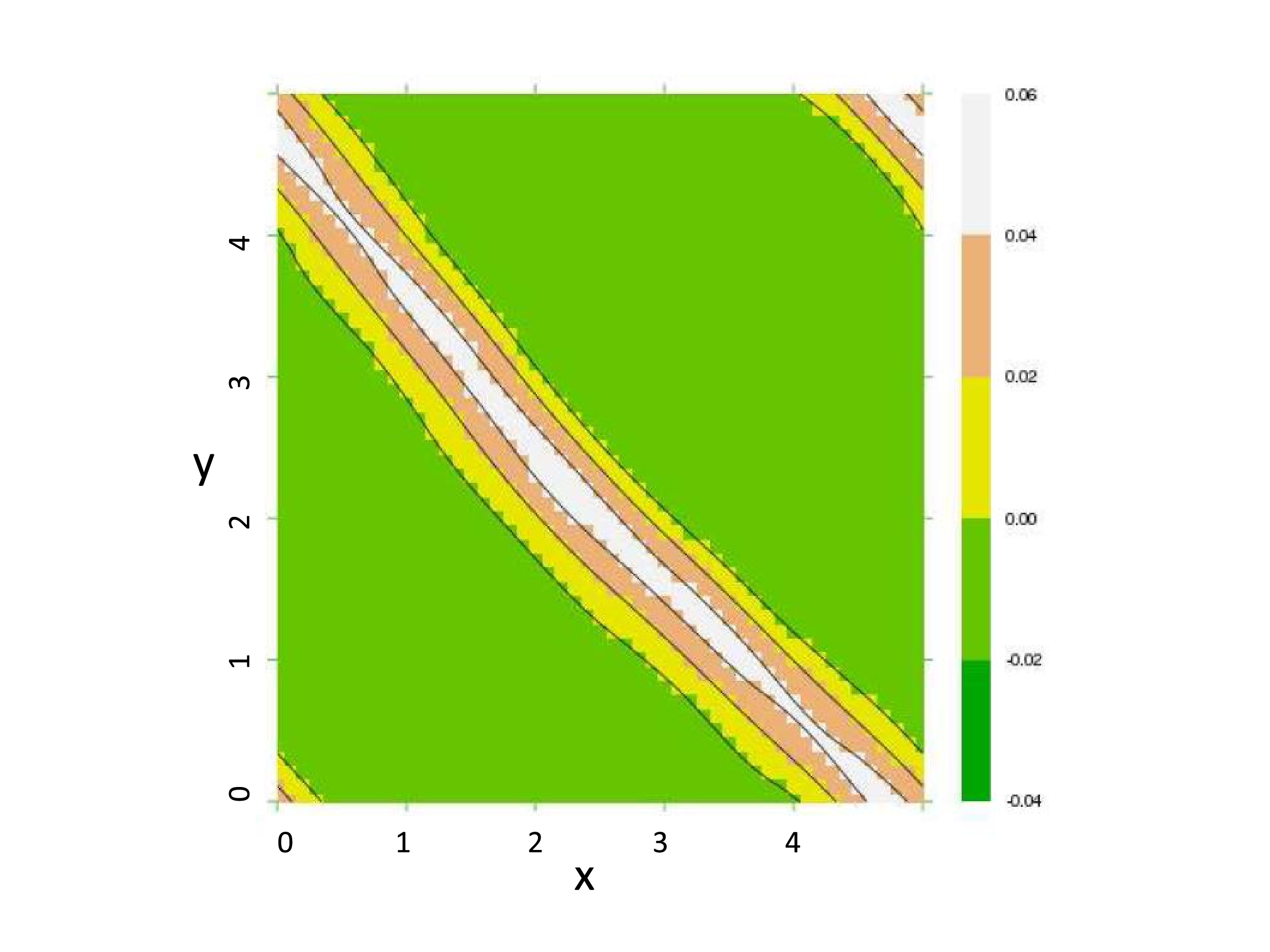}\hskip 0.7cm \includegraphics[scale=0.28]{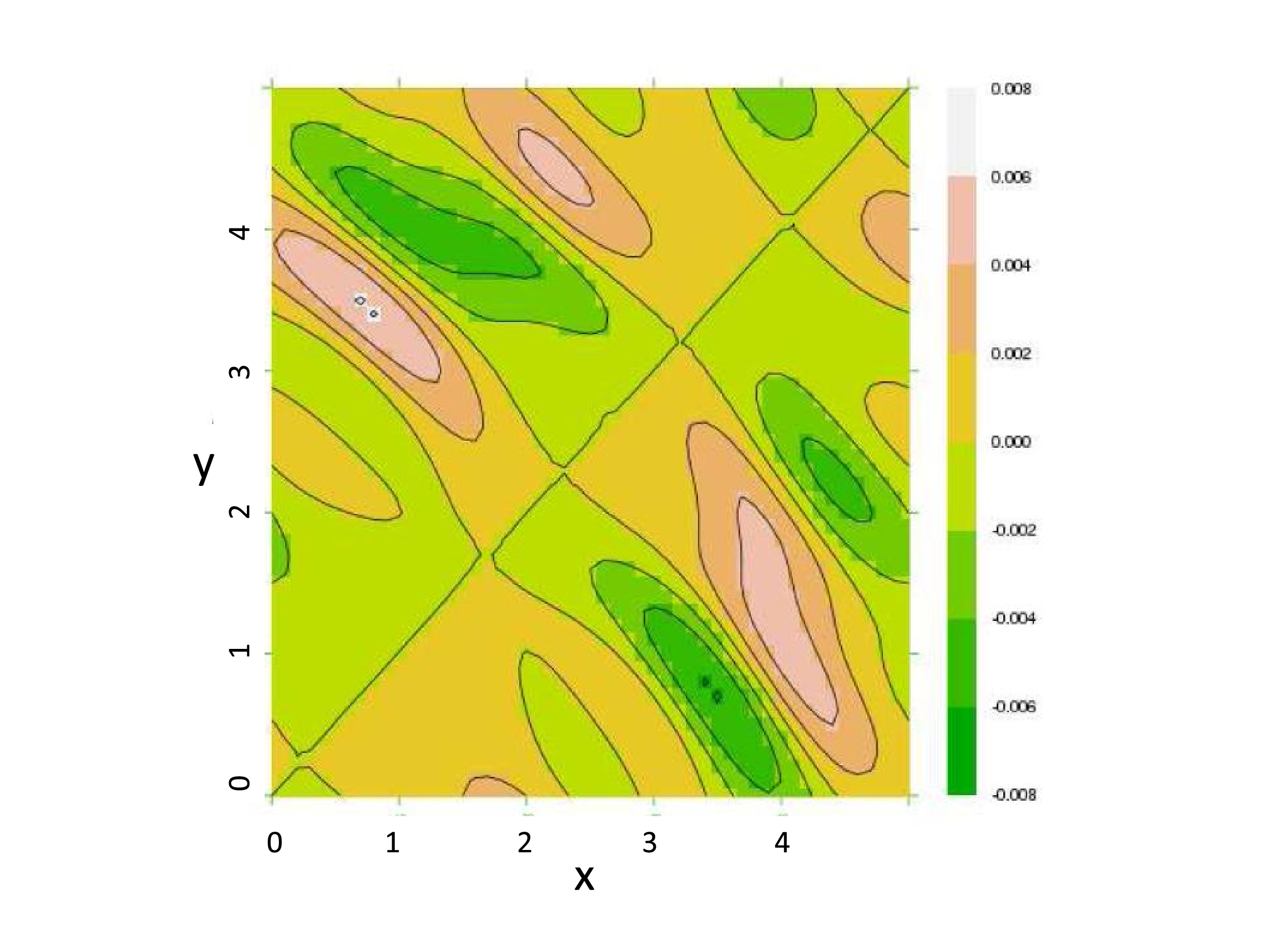}
   \caption{Level plots of the potentials $u(x,y)$ (left) and $w(x,y)$ (right), obtained by averaging over all stocks. Note the high ridge appearing for $u(x,y)$ for $x+y \approx 5$, and the complex flow pattern induced by $w$. These patterns are found to be very similar for all stocks and time periods.}
 \end{center}
  \end{figure}

These potentials are represented in Fig 9. Again, the patterns are very robust and appear to be
significant even in regions where the probability to find a queue is small (i.e. $x>3$ or $y>3$). Interestingly, and for reasons we do not understand, the potential $u(x,y)$ only depends, in 
a first approximation, on $r=x+y$; it has a broad, shallow minimum around $r \approx 2$ (corresponding to most probable queues with $x \approx y \approx 1$), a
sharp maximum around $r \approx 5$ (corresponding to the white ridge seen in Fig. 9-a), followed by a secondary minimum around $r \approx 7$. Note however that this secondary minimum 
does not lead to a peak in the stationary distribution $P_{st}(x,y)$ because the diffusion coefficients $d_x$ and $d_y$ are very large in this region (see Fig. 4). 
The rotational component of the drift, represented in Fig. 9-b, is quite complex, but its structure, in the most relevant region $x \sim y \sim 1$ can be summarized as follows:
close to the diagonal $x=y$, this component of the drift is towards zero, i.e. queues of similar size tend to shrink together. When one queue is much smaller than the other, the drift is
directed towards the diagonal (queues tend to equilibrate), before bending towards zero again closer to the diagonal. It would be very interesting to build a theory that could explain the intricate pattern displayed by the drift field $\vec f$, 
especially because, as emphasized above, this pattern appears to be stable across stocks and across time. The Mean-Field Game approach of \cite{CAL}, appears to be a way to approach the problem.

Note that the presence of the rotational component prevents us from writing down the Boltzmann-Gibbs measure that generalizes Eq. (\ref{GB}) to the two-dimensional case, even in the
region $x \sim y \sim 1$ where $d_x=d_y=$ constant. However, from the general pattern of the flow field shown in Fig. 8, it is intuitively clear that the resulting stationary distribution $P_{st}(x,y)$
should have the humped shape shown in Fig. 2-b.

Finally, the quantities $q_\pm(x|y)$ and $P_\pm(x,y)$ can be studied (not shown here). The noticeable patterns are: 
\begin{itemize}
\item when $x \sim y \sim 1$, the probabilities of price changing events $q_\pm(x|y)$ reach a minimum. $q_+(x|y)$ remains small as $x \to 0$, $y \sim 1$, which means that if the bid becomes much smaller than the ask, the probability that the bid goes up is small, which makes sense 
since the sell pressure on the ask is larger than the buy pressure on the bid. Conversely, as expected, $q_-(x|y)$ remains small when $x \sim 1$ but $y \to 0$; 
\item  $P_-(x,y)$ has a double peak structure: conditionally to an event where the best price disappears and the second best price takes over, the most probable size of the queue is $x \approx 1$, 
while the opposite queue either has a typical size ($y \sim 1$), or is relatively small ($y \ll 1$). However, as we noticed just above, the probability of these events is small. 
\item $P_+(x,y)$ has a sharp peak for $x \sim y \ll 1$, and a broader peak for $y \sim 1$ and $x \ll 1$ i.e. the newly created improved bid has a small volume (as expected), and the most 
probable situations are either that the old ask is small as well, or that it has a typical value. 
\end{itemize}

\section{Summary \& Conclusion}

Motivated by empirical data, we have proposed a statistical description of the queue dynamics for large tick assets based on a two-dimensional Fokker-Planck (diffusion) equation, that
explicitly includes {\it state dependence}, i.e. the fact that the drift and diffusion depends on the volume present on both sides of the spread. ``Jump'' events, corresponding 
to sudden changes of the best limit price, must also be included as birth-death terms in the Fokker-Planck equation. All quantities involved in the equation can be calibrated 
using high-frequency data on best quotes. One of our central finding, 
repeatedly emphasized throughout the paper, is the the dynamical process is {\it approximately scale invariant}, i.e., the only relevant variable is the ratio of the
current volume in the queue to its average value. While the latter shows intraday seasonalities and strong variability across stocks and time periods, the dynamics of 
the rescaled volumes is universal. In terms of rescaled volumes, we found that the drift has a complex two-dimensional structure, which is a sum of a gradient contribution and 
a rotational contribution, both stable across stocks and time. 
This drift term is entirely responsible for the dynamical correlations between the ask queue and the bid queue. The structure of the diffusion term, on the
other hand, is found to be quite trivial, with no dependence on the opposite volume. 

Although our scale invariance assumption is, we believe, a suitable first approximation to describe queue dynamics, a detailed study of the violations of this assumption would be
interesting and could reveal some systematic dependence on stock characteristics (price, liquidity, market cap, etc.) or hour of the day, for example. Clearly, scale invariance should only hold for sufficiently 
large volumes in the queues; we therefore expect that violations will be more pronounced for smaller average volumes and will be very strong for small tick stocks. 
Another issue that would certainly deserve further work is whether the universality uncovered here for NASDAQ stocks extends to other types of large tick securities with time priority 
(for example, non US stocks, large tick futures contracts, etc.)

Another open question is the validity of the Fokker-Planck (diffusion) framework, which amounts to truncate the Kramers-Moyal equation to second order. Such a truncation is 
not immediately justified since the distribution of elementary volume changes $\Delta V$ for each event (execution of a market order, addition or cancellation of a limit order) is
found to have heavy tails. Still we have shown that when solved to give the stationary distribution of rescaled volumes in a queue, the Fokker-Planck equation calibrated on dynamical data 
fares quite well at reproducing the empirical (static) distribution. 

It would also be very interesting to develop a theory, based on equilibrium, optimizing agents, or on agents using heuristic/behavioral rules, able to reproduce the fine details of 
the flow field $\vec f$ shown in Fig. 8. As mentioned above, the statistical determination of this flow field is quite accurate, and the pattern appears to be robust across stocks and 
time periods, once expressed in reduced volumes. We believe that this comparison will prove to be a stringent test for theoretical assumptions on the behaviour of agents in financial markets.

Finally, we want to emphasize that the theory developed above is not complete. For example, it does not allow us to answer a crucial question as far as optimal execution is concerned, i.e.: if I place
a sell order on a queue of volume $V_A$, knowing that the opposite volume is $V_B$, what is the probability that my order will be executed, and how long should I wait? The answer to
these questions require an additional information, absent from the above framework, which is the position in the queue of the cancelled orders. While added orders are always at the 
back of the queue, cancelled orders can be anywhere in the queue. Clearly, the position of these cancelled orders matter, and determine the speed at which my own order makes it to the top.
Our preliminary statistical analysis suggests that the probability $q(H|L)$ that the $H$-th order is cancelled, in a queue that contains a total of $L$ orders, again takes a scaling form:
$q(H|L) \propto {\cal Q}(u)$, where $u=(L-H)/L^{1/3}$ and ${\cal Q}(u)$ is a decreasing function of $u$. This means that, as expected, the orders most likely to be cancelled are those at the back of the queue -- this statement becoming sharp as the height of the queue $L$ goes to infinity. The unexpected finding, for which we have currently no interpretation, is that the width of the region where these orders are cancelled grows with the height of the queue as a fractional power, $L^{1/3}$. We leave this as an intriguing open question.

\section*{Acknowledgements} We want to thank Xavier Brokmann, Charles Lehalle, Marc Potters and Spyros Skouras for very helpful discussions and suggestions, Olivier Guedj 
who participated to the first stages of this study and Aurelien Vall\'ee for help setting up the order book animation.


\end{document}